\documentclass[runningheads]{llncs}
\usepackage{graphicx}
\usepackage{booktabs}
\usepackage{verbatim}
\usepackage[tableposition=above]{caption} % proper spacing of captions under tables
\usepackage{subcaption}
\usepackage{csquotes}
\usepackage{algorithm}
\usepackage{algpseudocode}
\usepackage{xcolor}
\usepackage[shortlabels,inline]{enumitem}

\usepackage{amsmath,amsfonts}
\usepackage[hyphens]{url}
\usepackage[hidelinks]{hyperref}
\hypersetup{breaklinks=true}

\usepackage{cite}   % to get multiple citations in numerical order
\usepackage[colorinlistoftodos]{todonotes}

\begin{document}

\title{A methodology for comparing and benchmarking quantum devices} %\thanks{Supported by organization x.}}
\titlerunning{Quantum Benchmarks}
\author{Jessica Park\inst{1,2} \and
Susan Stepney\inst{1}\orcidID{0000-0003-3146-5401} \and
Irene D'Amico\inst{2}
\email{\{jlp567,susan.stepney,irene.damico\}@york.ac.uk}
}
\authorrunning{J. Park, S. Stepney, and I. D'Amico}
\institute{Department of Computer Science, University of York, UK  \and
School of Physics, Engineering and Technology, University of York, UK
}

\maketitle              % typeset the header of the contribution
\begin{abstract}
Quantum Computing (QC) is undergoing a high rate of development, investment and research devoted to its improvement.
However, there is little consensus in the industry and wider literature as to what improvement might consist of beyond ambiguous statements of \enquote{more qubits} and \enquote{fewer errors}. 
Before one can decide how to improve something, it is first necessary to define the criteria for success: what are the metrics or statistics that are relevant to the problem? 
The lack of clarity surrounding this question 
%and its potential answers 
has led to a rapidly developing capability with little consistency or standards present across the board. 
This paper lays out a framework by which any user, developer or researcher can define, articulate and justify the success criteria and associated benchmarks that have been used to solve their problem or make their claim. 
%This clarity should also aid in the search for and proof of quantum supremacy, which has thus far seen a myriad of claim and counterclaims. 
%\keywords{Quantum computing  \and Benchmarking \and Fidelity}
\end{abstract}
\section{Introduction and Motivations}
The theory of quantum mechanics and its potential use in computing having been under research for over 100 years.
Quantum computing hardware is still in relative infancy, however, with limitations in both size (qubit number) and capability (error rates).
The current state of play is often referred to as the Noisy Intermediate Scale Quantum (NISQ) era. 

\textit{Noisy} refers to the fact that there are unprogrammed interactions present both between qubits in the device, and between  qubits and their environment; these interactions may cause the device to perform in a non-ideal way,
and be a source of errors. 
%This noise is more generally considered the source of errors seen in quantum computing devices. 
A description of the causes and effects of errors in quantum devices is presented in Sec \ref{sec:errors}.

\textit{Intermediate Scale} refers both to the limited number of qubits in these devices, and to the limited connectivity between them. 
%This is important because 
Desired (non-noisy) interactions between qubits require direct connection; 
%(or at least they are supposed to, see Sec \ref{sec:errors}) 
limited connectivity limits the circuits that can be implemented natively on the device. 
SWAP gates can be used to implement indirect connections, but at the expense of potentially more errors and longer circuit times. 

Being able to characterise both the scale and noise of a quantum device is crucial both for monitoring the development of the field and for determining if a particular device is fit for a particular purpose. 
There are a number of ways that quantum devices are being characterised at present, but there are no standard measures for comparing devices across manufacturers, qubit realisations or computational models. 
This is discussed further in Sec~\ref{sec:Bench}.

After discussing these issues (Sec~\ref{sec:theory}), we propose a methodology for designing and performing characterisation experiments on quantum computers (Sec \ref{sec:Method}),
and illustrate its use through an example of the design of a benchmarking experiment where the relevant choices are documented and justified (Sec~\ref{sec:Example}).
%Although we believe that this approach is widely applicable and effective, it does have some limitations and exclusions (Sec~\ref{sec:Exc}). 

\section{Background}\label{sec:theory}
\subsection{Errors in Quantum Computing}\label{sec:errors}
Any effects or interactions unaccounted for in one's model of the ideal hardware are equivalent to unknown terms in the actual Hamiltonian that is encoded and enacted on the device.
When the enacted Hamiltonian is not equal to the desired Hamiltonian, the energy landscape that it describes is also different. If the discrepancy is severe enough, or the landscape is particularly rugged, the two Hamiltonians may have different ground states. 
The ground state determines the solution of the problem, so different ground states means that the optimal solution cannot be found.  

When the mathematical form of the unintended interactions can be estimated, they can be included in the Hamiltonian to model their effects on the quantum solver.
Example effects include random noise, onsite energies, next-nearest neighbour interactions and inter-excitation interactions.
For example, the effects of these on perfect state transfer in spin chains have been modelled \cite{Ronke2011-rp}.
By understanding the relative effects of each type of phenomenon on the fidelity of state transfer, one can then focus on the elements of the hardware or software that may need to be optimised.  

D-Wave, who manufacture and sell quantum annealers, include in their documentation details on Integrated Control Errors (ICE), which limit the dynamic range of the user defined $h$ and $J$ values, the programmable values that encode the problem to be solved \cite{D-Wave_Systems_undated-yb}.
The five main sources they identify are: 
\begin{enumerate}
    \item Background Susceptibility: Next-nearest neighbour interactions and applied bias ($h$ value) leakage between neighbouring qubits
    \item Flux Noise of the Qubits: Noise that varies with time and logical qubit size (chain lengths)
    \item DAC Quantisation: Precision effects between digital and analogue controls 
    \item I/O System Effect: Parameter control depending on aspects such as annealing schedule
    \item Distribution of $h$ and $J$ Scale Across Qubits: Fabrication imperfections that give qubits slightly different properties. This is also a time-varying error
\end{enumerate}

A number of other errors sources are also mentioned in brief, such as temperature, photon flux and readout fidelity. 
The challenges that these effects present in real use cases for the D-Wave machines are well documented 
%Chancellor (2022), Nelson (2022) \& Pochart (2022) 
\cite{Pochart2022-gi, Chancellor2022-ya}.

%When considering the effect that any noise and errors may have on the device and therefore the results, one can do error modelling and estimation; the results can be used to aid decisions when selecting or using a device. 

Error modelling and estimation can be used to investigate the effect of noise and errors on a device and its results, and aid decisions when selecting or using a device. 
For example, algorithms such as EQUAL inject perturbations in the quantum Hamiltonian  to mitigate the estimated systematic error in the D-Wave quantum annealer \cite{Ayanzadeh2021-lj}. 
%With an understanding of the DAC quantisation errors found in D-Wave devices, this protocol  perturbs the Hamiltonian to shift to one energy level of the desired Hamiltonian. 
%Although the actual precision of the quantisation is not known, it is estimated through a trial and error approach and found to typically be not more than 8 bits of precision. 
%This approach considers only whole device level errors; heterogeneous component errors are not addressed. 

It is important to understand how the errors in the hardware affect the probability of success (defined as finding a state with lower energy than the known excited states of the intended problem) and how this scales with problem size.
Albash et al \cite{Albash2019-zr} study this in both simulated quantum annealers and the D-Wave 2000Q. 
%The difference between the simulated and real results presented there shows the effect that other errors in the annealer are having. The study does not address the origins of the noise or the exact modelling of the error conditions: it assumes that Gaussian noise with zero mean and variable standard deviation is an accurate representation of the `analog errors'. 
%The study suggests that repetition code, as in typical error-correction codes, could be used to create an acceptable level of fault tolerance in quantum annealers. 
The study considers the quantum computing chip to have one overall error and does not attempt to attribute it to any particular effect. 

Zaborniak and de Sousa \cite{Zaborniak2021-rz} begin to characterise the noise seen in both D-Wave2000Q chip and the newer 5000 qubit Advantage System.
Their results show that in the 2000Q system, the noise is frequency dependent according to \(1/f^{0.7}\), and in the Advantage system, the noise amplitude is 2--3 times larger whilst also being affected by further noise sources at short annealing times. 
The mathematical description and numerical results for the 2000Q system fit with the errors described by the D-Wave documentation when explaining the ICE.2-Flux Noise \cite{D-Wave_Systems_undated-yb}. 
This presents evidence that on-chip measurements can be directly related to theoretical sources of error, and that the newer, larger device is subject to larger errors. Further research on this is likely to shape the future development and usage of quantum annealing chips. 

Quantum computing architectures have  further parameters involved in the running of the solver that affect the quality of the solutions. 
D-Wave parameters are user controlled in the accompanying software \cite{D-Wave_Systems_undated-yb} and include spin reversal (to mitigate the effect of spin bath polarization) and anneal offsets (changing the qubit freeze-out rate). 
Barbosa et al \cite{Barbosa2021-mr} show that these parameters can be optimised for entire classes of problems, removing the need for computationally expensive optimisation for each individual problem.
%A fixed problem embedding is used throughout the experiment deliberately as an invariant when considering multiple problems from the same class. 
A fixed problem embedding is used when considering multiple problems from the same class.
This allows for easier attribution of the optimisation, but at the cost of potentially not fully optimising the device or preventing the optimisation from generalising. 
Each parameter is optimised independently; the authors state that
simultaneous optimisation would likely yield better results. 

Most of the error measurement schemes discussed in the literature involve either running specialised benchmarking algorithms \cite{Zaborniak2021-rz}, or consider outputs of large numbers of runs \cite{Albash2019-zr}. 
The first method is generally computationally expensive and the second ignores any time dynamics to the noise. 
A continuous monitoring scheme  that does not involve any additional algorithms \cite{Zolotarev2022-pf} has been tested on both emulated and real quantum hardware and on both single- and two-qubit gates.
Drawbacks include the limitation to small circuit sizes and inherent assumptions on the noise model; these are areas of future research. 

Pelofske et al \cite{Pelofske2023-jc}
suggest another approach to continuously monitor performance of a quantum annealer. 
The basic principle is to embed a second QUBO, the Performance Indicator (PI),  onto the quantum annealer using  unused qubits that remain after the problem QUBO has been embedded. 
%This second QUBO is called the Performance Indicator (PI).
The set-up phase involves running multiple anneals and subsequent readouts of the PI and problem QUBO, and creating a time series of the moving averages. 
Since the PI has comparable trends in performance to the problem QUBO, the PI performance can be used to predict the quality of the result from the problem QUBO. 

In addition to whole-chip measurements, one can  consider computing fidelity at the single qubit level. 
Nelson et al \cite{Nelson2021-ca} have developed Quantum Annealing Single-qubit Assessment (QASA)  to characterise individual qubits within a quantum annealer. 
Each qubit is repeatedly sampled through a range of input fields, and four parameter values are extracted: effective temperature, noise, transverse field gain, and bias. The authors are clear that, although their model and associated parameters appear to fit the results of the measurements, it does not account for the physical origins of the behaviours. 
The QASA protocol is performed for all the qubits within a chip in parallel, and the variations and correlations across the chip  analysed. 
There are spatial variations across the chip, and there is a difference in behaviour between the horizontally and vertically oriented qubits. 
Their results show clear heterogeneity at the very lowest level of hardware. 
Although the QASA protocol is aimed at the single qubit level, the authors have also considered how changes in other layers of the model may affect their results by repeating the experiment with different annealing schedules. They show that bias and transverse field gain are largely robust to a change in anneal schedule, but effective temperature increases logarithmically with anneal time. 
Hence, when optimising or correcting for effective temperature, the anneal time needs to be taken into account. Their results show the importance of full stack considerations, from detailed hardware to abstract algorithms, when developing a quantum computing pipeline.

\subsection{Error Correction and Mitigation}
Because of the high levels of noise and errors in NISQ devices, quantum error correction and mitigation methods are needed to make best use of these machines.
Extensive research into the theory and measurements of errors in these devices has lead to increasingly effective error correction schemes.
We overview a selection of these methods,
chosen to highlight their range. 

The quantum fault-tolerance theorem states that, given uncorrelated noise and a sufficiently low gate fail rate, a quantum computer can give arbitrarily accurate solutions to computations of any length \cite{Pearson2019-xz}. 
In quantum gate model computers, error correcting protocols can be used to achieve the desired accuracy with some qubit and gate overhead \cite{Aharonov2008-ow}. 
One theory of quantum error correcting codes,  by Knill and Laflamme \cite{Knill1997-hp}, is based on the idea of \enquote{encoding states into larger Hilbert spaces subject to known interactions}. 
The larger Hilbert space is made of ancilla qubits, and the exact encoding is designed such that any gate fault preserves the quantum information in the state. 
There are a number of different quantum error correction algorithms that are based on encoding one logical qubit in anywhere between 5 and 9 physical qubits. 
Quantum error correction codes work well in theory with gate model computers, and give the potential for universal quantum computation. 
However, they reduce the number of logical qubits achievable with a given device.

Quantum error correction codes are typically discrete in nature, using unitary gates to flip states to recover lost information. 
In practice these are difficult to implement in a number of qubit types including SQUIDs \cite{Kelly2015-pg} and trapped ions \cite{Hennrich2011-ca}. 
Ahn et al \cite{Ahn2002-pr}
proposed a scheme of continuous measurements in situations where discrete error correction is not applicable; Hamiltonian operators filter out the noisy signals. 
Borah et al  \cite{Borah2022-gc}
have implemented this scheme, and shown it provides fidelity improvement over uncorrected states. The parameter that controls the feedback strength is important in preserving state fidelity: feedback and detailed knowledge of one's system is required for optimisation. 

Developing error-correction protocols for quantum annealers is an area of active research. 
It is as yet unproven whether a noisy quantum annealer could accurately replicate an ideal device for arbitrarily large problems. 
However, there are proposed methods for error correction in annealing qubits. 
Inspired by circuit model error correction,  Pudenz et al \cite{Pudenz2014-oy} use ancilla qubits to create repetition; this imposes an energy penalty to stabilise the bit flip operations across the ancilla qubits of the same logical qubit. 
They demonstrate 344 physical qubits emulating antiferromagnetic chains of lengths between 0 and 90 spins, and find improved performance over both classical and uncorrected quantum methods. 
These results are promising; however, the parameters that define the energy penalty have to be tuned to the specific problem, which requires some advance knowledge of the correct solution. 
Further work is needed to use this protocol more universally.
%, including for problems in which the required solution is not known \textit{a priori}. 
According to the methodology proposed below,  energy penalty parameters form part of the formulation stage in the information processing pipeline, and could therefore be included in the overall optimisation algorithm. 

Classical post-processing is often used for error mitigation schemes. 
There are a number of different schemes,
%for this at various levels of performance improvement and complexity, 
ranging from bit-flip hill climbing (that ignores spin interactions) to randomised multi-qubit correction (RMQC) in which shuffled groups of quits are flipped in such a way that tunnels through to lower energy states of the whole system \cite{Ayanzadeh2021-zo}. 
%RMQC works on the assumption that each sample from the annealer output has some components of the desired ground state (i.e. the errors are not extreme enough to completely separate the low energy levels of the desired and enacted Hamiltonians). 
%From the testing carried out in the development of RMQC, this assumption is found to be valid, although many of the test problems are robust and do not require any error correction. 
A benefit of the RMQC approach is that it is deterministic (unlike most post-processing methods) and does not require hyperparameter tuning, so there are fewer variables to optimise.

\subsection{Device Characterisation and Benchmarking} \label{sec:Bench}
Measuring the errors in a quantum device is a way of characterising that device.
This provides useful information for manufacturers and researchers, 
but it is not simple for users to discern how errors might affect performance when applied to their real world problem. 
%This is confounded when error mitigation schemes could and are being used either deliberately by the user or within the calibration or `default' usage of the device as provider by the manufacturer. 
It is not always clear in benchmarking literature whether error mitigation schemes have been used, and if so, how that has affected not only the results of the benchmark but also the complexity in number of qubits and circuit depth. 
If the error mitigation happens by default when using a particular device, we have to rely on the provider to complete these steps successfully and reliably. When manufacturers change their default processes over time, these ned to be communicated to users, so that they can be taken into account when comparing past results.  

The majority of the leading quantum device companies produce gate model devices; key exceptions are D-Wave (annealers) and a handful of photonic specialists (PsiQuantum, Xanadu and ORCA).
%Due to the heterogeneity available to developers of quantum computing hardware, 
It can be  difficult to compare performance between different hardware architectures and qubit types. 
Even properties that can be attributed both to circuit model devices and to annealers, such as number of qubits, cannot be compared easily: they affect the operation very differently. 
Including photonic and topological devices makes any comparison  harder still. 
A particular challenge when comparing commercial quantum solvers is lack of consistency in reporting fidelity or error measurements. 
As this is a new field, there is no precedent for what measures users need, understand or care about when choosing a device. What measures are important depends on specific use cases and algorithms.  
In many cases, users access the devices indirectly through cloud software,
%such as Amazon Web Services BraKet service. 
and it is infeasible for users to measure the relevant values, due to high calculation costs and the amount of device control required. 
With more discerning users with more pressing problems, providers will need to conform to a standard set of performance measures; but which ones?

Suau et al \cite{Suau2021-wp} have measured
some operational parameters such as qubit response, bias and saturation in both major hardware architectures. 
The authors recognise that there are more thorough qubit analyses that can be done in either circuit model or annealers, but state that the benefit of their approach is its applicability to a variety of hardware types. 
Their work raises an important issue:
%in the further development of quantum computers: 
in order to compare approaches, it is necessary to have some comparison measures. 
This issue gets increasingly fraught as hardware setups include different architectures and qubit types at various granularities within the same device. 

Quantum Volume \cite{Moll2018-vu}
%, $\tilde{V}_Q$, 
is one candidate measure. 
It 
%was first proposed by Moll \cite{Moll2018-vu} and 
is defined as
%\begin{equation}
	$\tilde{V}_Q = min[N, d(N)]^2$,
%    \label{eq:QV}
%\end{equation}
where $N$ is the number of qubits, and $d$ is the circuit depth, the maximum number of steps (unitary operations) executed in one run of the device. 
%One step is a unitary operation that can consist of a combinations of qubit gates that occur in parallel.
Different connectivities and error rates limit how many steps can be completed. 
Variants consider the largest square circuit ($N=d$) that a quantum computer can implement  \cite{Cross2019-pj}. 

Pelofske et al \cite{Pelofske2022-dy} compare the quantum volume of 24 NISQ devices  available to users in 2022. 
The quantum volumes are determined in-house, using the same approach, rather than taking values determined by the manufacturers or previous investigations. 
%What is important to note here is that 
The particular transpilation/compilation methods used by each device have an effect on the maximum circuit size that can be executed, and therefore on the resulting quantum volume. 
So users need to consider how a device is to be used, and its compatibility with the desired problem(s), when choosing a suitable device.

Quantum volume appears to be a useful measure for comparing circuit model quantum computers.
It is unclear how this might translate to either annealing or topological quantum computing architectures, however, where $N$ has  different implications  and $d$ is not always well-defined.  
Some attempts have been made to quantify the Quantum Volume of D-Wave quantum annealers by using equivalent problems that it can solve \cite{Tiziano2020-wo}; since quantum annealers are not thought to be universal solvers, the applicability of this approach is not clear. 

There are a number of quantum computing benchmarks focused on performance against particular problems. 
These are advertised as more useful measures for the user than system parameters such as coherence time or gate fidelity, and can provide a more direct comparison between fundamentally different qubit and architecture types \cite{Lubinski2021-hl}.
Benchmarking tasks can typically be applied to a range of devices or parameter scenarios, to provide information that can  be generalised to performance on problems of interest to the user. 
Atos (a digital transformation company) have developed a Q Score \cite{Martiel2021-ii}, a single number measure of the largest number of variables in a `Max Cut' problem that a device can optimise. 
%Atos have made the software freely available and invite manufacturers to self-assess against this measure. 
There are other similar application-driven measures developed by a number of companies 
%The Boston Consulting Group have put together a table comparing a number of device comparison measures ranging from system-driven to application-driven 
\cite{Langione2022-ho}.
%This can be seen in Figure \ref{fig:BCG}. 

How the performance of a benchmark task translates to the real world problems that a user is interested in determines the  utility of this kind of measure. 
%\notess{I have dropped figure 1 to save space (we'd need copyright permission to reproduce it anyway, I think?) - Happy to drop the figure}
% \begin{figure}
%     \includegraphics[scale=0.75]{BCG.png}
%     \caption{\small A comparison on benchmarking methods for quantum computing devices as produced by Boston Consulting Group \cite{Langione2022-ho}.
%     }
%     \label{fig:BCG}
% \end{figure}
It may not be practical or useful to compare  quantum hardware in isolation from the specific problem set of interest. 
It might be that, for each problem complexity and class, different quantum computers should be compared only on the quality of- and time to- solution, themselves complex functions of qubit number, error rate, connectivity etc.
Linke et al \cite{Linke2017-ii} compare two 5 qubit systems with different qubit realisations and connectivities, finding that the architecture that most closely matches the desired circuit tends to perform better. 

These kinds of results, albeit on a small scale, suggest that different quantum processors should be chosen for different tasks or even subtasks within a processing pipeline; but how to choose?

\section{Proposed Methodology} \label{sec:Method}
There are a great number of choices that have to be made when characterising and comparing quantum devices. 
Even for users of the devices, who may not be making these choices themselves, it is crucial that the decisions that have been made are well communicated and justified for the measure to be useful. 

We propose the following methodology for deciding on and executing any type of characterisation and benchmarking, designed to provoke a deep consideration of the desired and unintended consequence of all the choices being made at each step. The steps are:
\begin{enumerate*}[label=(\arabic*)]
    \item Purpose
    \item Success Measure
    \item Generalisability
    \item Robustness
    \item Expressivity
    \item Limitations,
\end{enumerate*}
discussed below.

%Each of these headers will now be expanded on in turn. The following section (\ref{sec:PurpSucc}) considers the first two points of defining your purpose and success metric. Then, section \ref{sec:Cons} discusses the remaining considerations that need to be main based on the first two points.

\subsection{Purpose and Success Measure} \label{sec:PurpSucc}
\subsubsection{Q1:} What is the main purpose of the intended measure? 
\begin{enumerate}[(a)]
    \item to compare  architectures and implementations of  quantum computers
    \item to compare  fabrications/alternative chips of the same architecture 
    \item to compare the relative difficulty of particular problems
    \item to monitor the effect of error correction/mitigation and calibration methods
    \item to measure the improvements and developments between iterations of quantum devices
    \item other
\end{enumerate}

When answering Q1, there is unlikely to be a single option that covers the purpose, but it is important to prioritise the different purposes, and to keep the decision in mind when answering Q2. 

\subsubsection{Q2:} How do you define \enquote{success} when considering a quantum computer? 
\begin{enumerate}[(a)]
    \item has the best of a desired hardware measure (eg. number of qubits)
    \item executes the process it was programmed to do 
    \item produces a quantum state that matches that from an ideal exact simulation
    \item produces a probability distribution that matches that from an ideal  exact simulation
    \item provides a \enquote{useful} result to a problem that is challenging to solve by other means (either due to solution quality or time to solution)
    \item other
\end{enumerate}

\noindent
\textbf{Objective function.}
The answers to these two questions give a purpose and definition for success.
These now need to be translated into an \textit{objective function} that gives a number or set of numbers that can be used to directly answer the question of purpose. 
The development of objective functions has been considered in length in the field of genetic algorithms (see Pare et al \cite{Pare2015-tt} for an image processing example).
Obtaining a useful objective function is key in ensuring the benchmark provides as much relevant information as possible. 
%For example, if a company is changing the manufacturing process of a particular product to improve quality and boost sales, a good objective function might include the profit margin on the product, the sales figures over some time window and the user ratings or return customer percentage. 
%These factors could then be combined in some linear weighting scheme or form different axes of a larger behaviour space to be explored.
Choosing an objective function is likely to be a trade-off between simplicity, interpretability, and multifaceted complexity.
%\notess{sentence sorts of peters out... I've tried to explain what I mean a bit more here}

\subsection{Generalisability, Robustness, Expressivity} \label{sec:Cons}
Given the answers to Q1 and Q2, and an objective function, there should now be a developing idea of a potential benchmark. Related to the options above, some appropriate choices might be:
\begin{enumerate*}[(a)]
    \item read off the device specification or get recent calibration data;
    \item Quantum Process Tomography to give process fidelity or Direct Characterisation of Quantum Dynamics;
    \item Quantum State Tomography to calculate trace distance between states;
    \item Total Variance Distance to calculate the similarity between distributions \cite{Paltenghi2022-im};
    \item compare result to an existing solution given by best available alternative device.
\end{enumerate*}
Although this is a small subset of the tools that could be used to benchmark, discussing them further highlights the importance of methodology considerations (3)--(5). 
Generalisability, robustness and expressivity are criteria by which the designer can consider whether the proposed benchmark is fit for (their specific) purpose. 
These criteria also provide a framework by which any documentation of the benchmark can justify the choices made. 

\textbf{Generalisability.}
Option (e) is the most overtly problem specific, yet any method chosen at this stage leads to further choices, and these affect the {generalisability} of the benchmark. 
This is mentioned briefly in  Sec.~\ref{sec:theory}, where particular embeddings are used to optimise a different area of the process.
In all options, one choice is the actual task to be given to the quantum device.
This typically means designing a circuit to complete a quantum process, prepare a quantum state, or provide a probability distribution. 
In the most general case, random circuits composed from the device's basis gates can be generated \cite{Boixo2018-nz, Cross2019-pj}; if enough circuits are run, we may assume the results to be maximally general. 
%This type of benchmarking is well documented in the literature \cite{Boixo2018-nz, Cross2019-pj, Hashim2022-rj}. 
From this, we here define \textit{generalisablity} as the ability for a quantum computer to complete a large range of tasks to a set standard.
Despite being maximally general in theory, Proctor et al \cite{Proctor2022-ve} show that randomised circuits are often limited to small numbers of qubits. 
By choosing a particular class of circuit (in that case, randomised mirror circuits), generalisability can be preserved whilst providing more opportunity to scale. 
Additionally, because those circuits have some structure, they are more similar to algorithmic circuits; this should make them more predictive of the device's performance on real world problems. 

\textbf{Robustness.} 
%Choices may affect the {robustness} of success measure. 
We define \textit{robustness} to be the stability of the performance to changes in the choices made during the benchmarking. 
In any quantum device, there are many operational parameters that affect  performance; when benchmarking, one must understand how these are used. 
Once a circuit has been designed, it needs to be embedded onto the hardware, taking into account the limited connectivity of the qubits. 
Most hardware manufacturers provide a tool that does this automatically for their devices in an efficient way, typically to minimise the number of additional SWAP gates.
Different providers, as well as having different connectivities, may have different algorithms that perform this step. 
%This is a good example of a potentially unconscious choice that is being made when using a particular benchmark, and opens up the questions in robustness.
%For example, 
How do the benchmarking results change if the choice of embedding algorithm is changed? 

Another operational parameter choice is the use of error correction or mitigation strategies in the benchmark task. 
If the top priority answer to Q1 is to measure the effects of these strategies, then this becomes the independent variable in the benchmark experiment. 
In other cases, the use of error reduction strategies should be considered and justified based on the purpose of the measure.
One device may exhibit better native performance; another device may be more receptive to error correction and therefore display overall better performance when these strategies are applied. 
Traditionally, benchmarks are considered useful for comparing so-called like-for-like performance; however, if the definition of a successful quantum computer is one that provides the most accurate result to a problem, then it might be more beneficial to compare best-for-best performance. 
Conversely, if the definition of a successful quantum computer also includes the time to solution, how these factors are balanced will influence any decision on the best benchmarking methodology.

\textbf{Expressivity.}
Another factor to consider when deciding on a benchmark is its ability to show significant difference between the results of different tests. 
When comparing quantum devices, one would like to be able to distinguish between potentially incremental improvements. 
We define  \textit{expressivity} as this range of results, complementing generalisability, which considers a range of tasks.
%\notess{Move this defn of generalisability to the relevant paragraph -- Done (line 363) and just recapped above}
For example, if the problem is too easy, all quantum devices may give a performance above 95\% (in some particular measure); it can seem as if the devices perform very similarly. 
However, when applied to a harder problem, these devices could show a great range in results.
An example of this is the MNIST dataset of handwritten digits, often used to benchmark machine learning classification algorithms, which  has been heavily criticised for its simplicity and lack of variability \cite{Hargreaves2020-rm}, with the PaperWithCode.com leaderboard (as of 28/04/2023) showing 63 algorithms with an error rate below 1\% \cite{noauthor_undated-mz}. 

Martiel et al  \cite{Martiel2021-ii}, when benchmarking quantum processors, define a $\beta^*$ function that defines how demanding a problem is; $\beta^*=0$ means the problem can be solved via a coin toss, and $\beta^*=1$ requires an exact solver. 
Their objective function for the benchmark then computes a score which can be compared to this threshold value for each problem. 

\subsection{Practical limitations} \label{sec:Exc}
It might not be possible to execute the full benchmark suggested by the methodology at this point.
%One of the key factors not addressed by the proposed pipeline is any practical limitations of being able to execute the chosen benchmark. 
First is the matter of cost:
executing a benchmark requires
time, access to the device, computational power,
and money.  These costs might make full benchmark execution too expensive. 
%As most users and researchers are accessing devices via cloud compute services, when tasks are computationally expensive (in terms of resources and time), this may also turn out to be monetarily expensive, which could be prohibitive to some. 
Second,
where ideal simulation results are required, this limits the size of the benchmark problems; quantum simulation on classical devices is currently limited to 10s of qubits for most tasks. 

%These are clearly important factors when designing a benchmarking algorithm and 
When faced with these limitations, the user needs to justify any reductions made to the ideal benchmarking experiment, and document how these limit the conclusions. 

\subsection{Justificiation and documentation of choices}
In order to make any benchmarking operation as informative and useful as possible, it is crucial that all the choices made in the process are documented and where possible, justified in way that relates back to the answers of the two key questions. 
In this documentation, any potential consequences of the choices made should also be discussed. 
Benchmarking studies are most useful when they are as transparent as possible in their execution and consistency is upheld wherever possible despite the heterogeneity in the range of quantum architectures, qubit realisations and potential algorithms. 

\section{Example Workflow}\label{sec:Example}
To demonstrate how the methodology proposed here might be put into practice, we provide an example here. 
%\notess{would this be better in 2nd person? - Changed, see below}

\textbf{Scenario:}
You are developing a quantum algorithm to perform image classification for medical screening purposes, initially on current NISQ devices, but more importantly for larger, upcoming models. 
You have tested  one quantum device, but before progressing further you would like to determine which device might give the best performance for your algorithm. 
You know that you require a gate model device, but have no other restrictions on access, qubit realisation etc.
What can you do in order to determine which quantum device to use for your application? 

\textbf{Q1: Purpose}.
From the scenario, your purpose is to compare different architectures and implementations of quantum computers (option (a) as listed above).  
Thinking a little deeper, a secondary purpose might be to compare alternative chips of the same architecture (option (b)) as it would be useful to know results are portable between chips of the same design. 

\textbf{Q2: Success Measure}.
%For question 2, I am asked to define success. 
In this scenario, success is most closely defined as option (e); providing a \enquote{useful} result to a problem. 
For this use case, you are more concerned with solution quality than time to solution. In image classification, performance is typically measured by properties like precision and recall, so it is likely that these will be factors in the objective function. 
An additional consideration here is option (b), that it executes the process it was programmed to do. 
Medical imaging is a high stakes problem, and you would like some assurance that on a gate level, the device is performing as expected.

From these answers to Q1 and Q2, it is now possible to define an objective function. 
In this case, you choose to define a simple objective function with a single parameter. 
You use the F1 score, the harmonic mean of precision and recall for the `true'  class \cite{Allwright2022-gg}. 
It is typically used in cases where there is an imbalance in the number of true and false cases. 
This is likely to be the case in a medical screening of a large population for a relatively rare condition. 

You could also choose a more complex two-dimensional objective function where the first dimension is the F1 score, 
and second dimension is a measure of how well the device in question can perform single-qubit rotation gates: $R_{x,y,z}(\theta)$.  
These gates are integral to running quantum neural networks, with the $\theta$ parameter  updated in the learning process \cite{Li2022-ok}. 
Having some assurance that these gates can be executed effectively provides confidence in the overall algorithm. 

The next step is to choose the benchmarking task. 
In this scenario, the task is the algorithm that you have already developed. 
Due to the focused task and use case, you already have a training and test set of labelled data. 
If this were not the case, or if the images in the data set are too large for the currently available quantum devices, you would have to choose a surrogate data set to use in the benchmarking. 
%\notess{Is the "too large" a show-stopper -- that would mean the devices couldn't execute your algorithm even once you'd benchmarked?? I think I just need to be clear here that the algorithm I'm developing maybe isn't for immediate use but to utilise larger devices as they become available. I've added some words in the initial scenario to this affect}

%I now move on to the second phase of the methodology by considering \textbf{generalisability}, \textbf{robustness} and \textbf{expressivity}. 

\textbf{Generalisability.}
You are not concerned with how well your dataset and subsequent results generalise to other problems. 
Generalisability in this case refers to choosing a dataset that represents your eventual (but currently unavailable) problem set. 
To do this, you need to consider the salient features of your dataset. 
Assume that the relevant medical images are typically in 8-bit greyscale with $2000 \times 2000$ pixels, and approximately 1 in 10 are labelled as true. 
The real images are too large so a dataset with smaller images should be considered.
Aside from this factor, you should try to find a dataset with 8-bit greyscale pixels in a square arrangement whereby the false case is present approximately 9 times more often than the true. 
To ensure generalisability, if time permits, you could even find multiple datasets that fit the criteria. 
%\notess{What's the point of benchmarking the application on smaller data, if your purpose it to really deploy on full size data (you are replicating Pathfinder, but we had a different purpose?) I think this is now covered in the clarified scenario of designing an algorithm that will be scalable for future devices.}

\textbf{Robustness.}
To consider robustness, you need to think closely about both the algorithm and the surrounding software used to access the quantum device. 
The algorithm requires embedding on to the quantum device, and the current process has been tuned to your existing quantum device. 
Because you want to test the devices and not the algorithm, you decide to use each manufacturer's recommended embedding tool for each device. 
In this way, you are comparing the best case (or at least the best-known) scenario for each device. 

There are other choices and predefined parameters that are used in the algorithm, such as the learning rate for the neural network and the method for entangling the qubits. 
In this case, you decide to use the best known values from the initial work, and keep these constant across all the tested devices. 
You have determined that these are unlikely to affect the comparative performance of the devices at this stage, and leave further refinement for latter stages. 

\textbf{Expressivity.}
%Expressivity is important when multiple devices could be expected to give a similar result, such as when the task is \enquote{too easy}. 
The difficulty of an image classification task can be increased by adding variation in the training set, such as by introducing edge cases, or adding artificial noise to existing images. 
Larger images are likely to provide a harder problem, due to a higher potential for natural variation, as well as requiring more qubits to be operational, increasing the need for more SWAP gates and also likelihood of gate errors from crosstalk. 
Where datasets are available publicly or have been used in previous studies, an impression of their potential expressivity can be gained, otherwise a good approach may be to have multiple benchmarks of different estimated difficulties. 

\textbf{Chosen Benchmark.} From considering these three factors, you are now able to choose an appropriate benchmarking task that is likely to achieve the aims set out in the answers to questions 1 and 2. 
In the scenario presented here, you choose 8-bit greyscale images that are 100--200 pixels in each dimension. 
You require two classes, where one class is 9 times more prevalent than the other; this could be achieved by down-selecting from a larger dataset. 
An example that fits these criteria is the publicly available \enquote{30K Cats and Dogs}, which has images that are in 8-bit greyscale and $150 \times 150$ pixels in size \cite{Unmoved2023-hj}.  
The full dataset contains around 15,000 of each class so for this experiment, I use 1,500 cats and 13,500 dogs. 

\textbf{Limitations.}
Due to the size of current NISQ devices, you have chosen a surrogate dataset for the benchmarking task. 
Although you have tried to ensure that it representative, there may sill be performance differences from the real case. The effects of this could be further limited by repeating the benchmark with a different dataset (with similar characteristics) and looking at how performance varies.
Another limitation that will affect your benchmarking are access to the quantum compute time and the wall-clock time you have available to complete the benchmark.
%Where these a deemed to have limited the utility of the benchmark, this should be properly documented.

One of the main benefits of using this methodology to determine a benchmarking strategy is that it encourages an awareness of all the intended and unintended consequences of the choices made. 
When one has this awareness, it is much easier to ensure that any literature written on the benchmarking task has adequate documentation and justification. 
Transparent benchmarking is important both within a development team and in sharing results with the wider community.

\section{Summary and Conclusion}
%We begin by laying out the motivation and theory behind running benchmarking tasks in the research and development of quantum computers.

%In light of the issues discussed, 
We propose a methodology for designing, running and documenting benchmarking tasks.
The aim is to ensure that tasks are chosen based on their intended purpose, rather than the alternative approach of completing `standard' benchmarking tasks and then attempting to determine what useful information can be gleaned from the results. 

The methodology begins by asking the user two questions to clarify their purpose, and their definition of success. 
Starting with these two focused questions, the user is then lead to develop an objective function to quantitatively measure the success as defined by the answers. 
After this step, the user has to determine a task that provides the data fed into the objective function.
We offer suggestions of what these tasks might look like depending on the objective function. 
When determining the benchmarking task, the methodology suggests three key considerations: generalisability, robustness and expressivity.
These factors have to be balanced depending on priorities brought to focus by the initial questions. 

As well as being a practical guide to determining the benchmarking task, the methodology also aids in the documentation and justification of the choices made,
%in this determination, 
and leads the user to think more deeply about the consequences of these choices. 
This represents a step towards transparency and coherency in a relatively young and rapidly developing field.

\subsection*{Acknowledgements}
The authors acknowledge Defence Science Technical Laboratory (Dstl) who are funding this research. 
Content includes material subject to © Crown copyright (2024), Dstl. This material is licensed under the terms of the Open Government Licence except where otherwise stated. To view this licence, visit \url{http://www.nationalarchives.gov.uk/doc/open-government-licence/version/3}, or write to the Information Policy Team, The National Archives, Kew, London TW9 4DU, or email: psi@nationalarchives.gov.uk

\Urlmuskip=0mu plus 1mu
\bibliography{paperpile}

\end{document}